\RequirePackage{lineno}	
\documentclass[a4paper,11pt]{article}
\usepackage{pos}
\usepackage[rightcaption]{sidecap}
\usepackage{wrapfig}
\newcommand{\pT}{\ensuremath{p_\mathrm{T}}}

\newcommand{\sNN}{$\sqrt {s_{\mathrm{NN}}}$~}

\newcommand{\DirPho}{$\rm \gamma_{\rm dir}$}
\newcommand{\gammaRich}{$\rm \gamma_{rich}$}
\newcommand{\piZro}{$\pi^{0}$}

\newcommand{\R}{$ R$}

\newcommand{\pTjet}{$p_{\mathrm{T,jet}}^{\mathrm{ch}}$}

\newcommand{\ET}{\ensuremath{E_\mathrm{T}^\mathrm{trig}}}
\newcommand{\IAAPYTHIASIX}{$ I^{\mathrm{\small{PYTHIA-6}}}_{\mathrm {AA}}$}
\newcommand{\IAAPYTHIAEIGHT}{$ I^{\mathrm{\small{PYTHIA-8}}}_{\mathrm{AA}}$}
\newcommand{\DeltaJetPt}{$\rm \Delta$\pTjet}

\newcommand{\RAA}{\ensuremath{R_\mathrm{AA}}}
\newcommand{\IAA}{\ensuremath{I_\mathrm{AA}}}
\title{Measurement of $\gamma$+jet and \piZro+jet in central Au+Au collisions at $\sqrt{s_{\mathrm{NN}}}$ = 200 GeV with the STAR experiment}

\author*[a]{Nihar Ranjan Sahoo (for the STAR Collaboration)}

\affiliation[a]{Shandong University,\\
Institute of Frontier and Interdisciplinary Science\\
 Qingdao, China}

\emailAdd{nihar@sdu.edu.cn, sahoo.niharr@gmail.com}

\abstract{We present the semi-inclusive measurement of charged jets recoiling from direct-photon and $\pi^{0}$ triggers in central Au+Au collisions at $\sqrt{s_{\mathrm {NN}}}$ = 200 GeV, using a dataset with integrated luminosity 13 $\mathrm {nb^{-1}}$ recorded by the STAR experiment in 2014. The photon and $\pi^{0}$ triggers are selected within transverse energy (\ET) between 9 GeV and 20 GeV. Charged jets are reconstructed with the anti-$k_{\mathrm{T}}$ algorithm with resolution parameters \R~=~0.2 and 0.5. A Mixed-Event technique developed previously by STAR is used to correct the recoil jet yield for uncorrelated background, enabling recoil jet measurements over a broad $\rm p_{T,jet}$ range. We report fully corrected charged-jet yields recoiling from direct-photon and $\pi^{0}$ triggers for the above two jet radii and also discuss the jet \R\ dependence of in-medium parton energy loss at the top RHIC energy.
}

\FullConference{%
  HardProbes2020\\
  1-6 June 2020\\
  Austin, Texas}

\begin{document}
\maketitle


Jet quenching arises from partonic interactions in the Quark-Gluon Plasma (QGP) formed in heavy-ion collisions~\cite{Wang:1994fx}. A valuable observable to probe the QGP is the coincidence of a reconstructed jet recoiling from a high transverse energy (high \ET) direct photon (\DirPho)~\cite{Wang:1996yh}, since \DirPho~does not interact strongly with the medium. A comparison of \DirPho+jet and \piZro+jet measurements may elucidate the color factor and path-length dependence of jet quenching~\cite{STAR:2016jdz}. In addition, a comparison of recoil jet distributions with different cone radii provides a probe of in-medium jet broadening. 

In these proceedings, we present the analysis of fully-corrected semi-inclusive distributions of charged jets recoiling from high-\ET\ \DirPho~ and \piZro~triggers in central Au+Au collisions at \sNN\ =~200 GeV. The data were recorded during the 2014 RHIC run with a trigger requiring an energy deposition greater than 5.6 GeV in a tower of the STAR Barrel Electromagnetic Calorimeter (BEMC), corresponding to an integrated luminosity of 13 $\rm nb^{-1}$. 
We compare the measured recoil jet yield in Au+Au collisions to a $pp$ reference via PYTHIA simulation and corresponding yield suppression is then further compared with theoretical calculations.
 We express the suppression in terms of jet energy loss and compare to other in-medium jet measurements at RHIC and the LHC.

\begin{wrapfigure}{l}{0.5\textwidth}
 \includegraphics[width=0.5\textwidth]{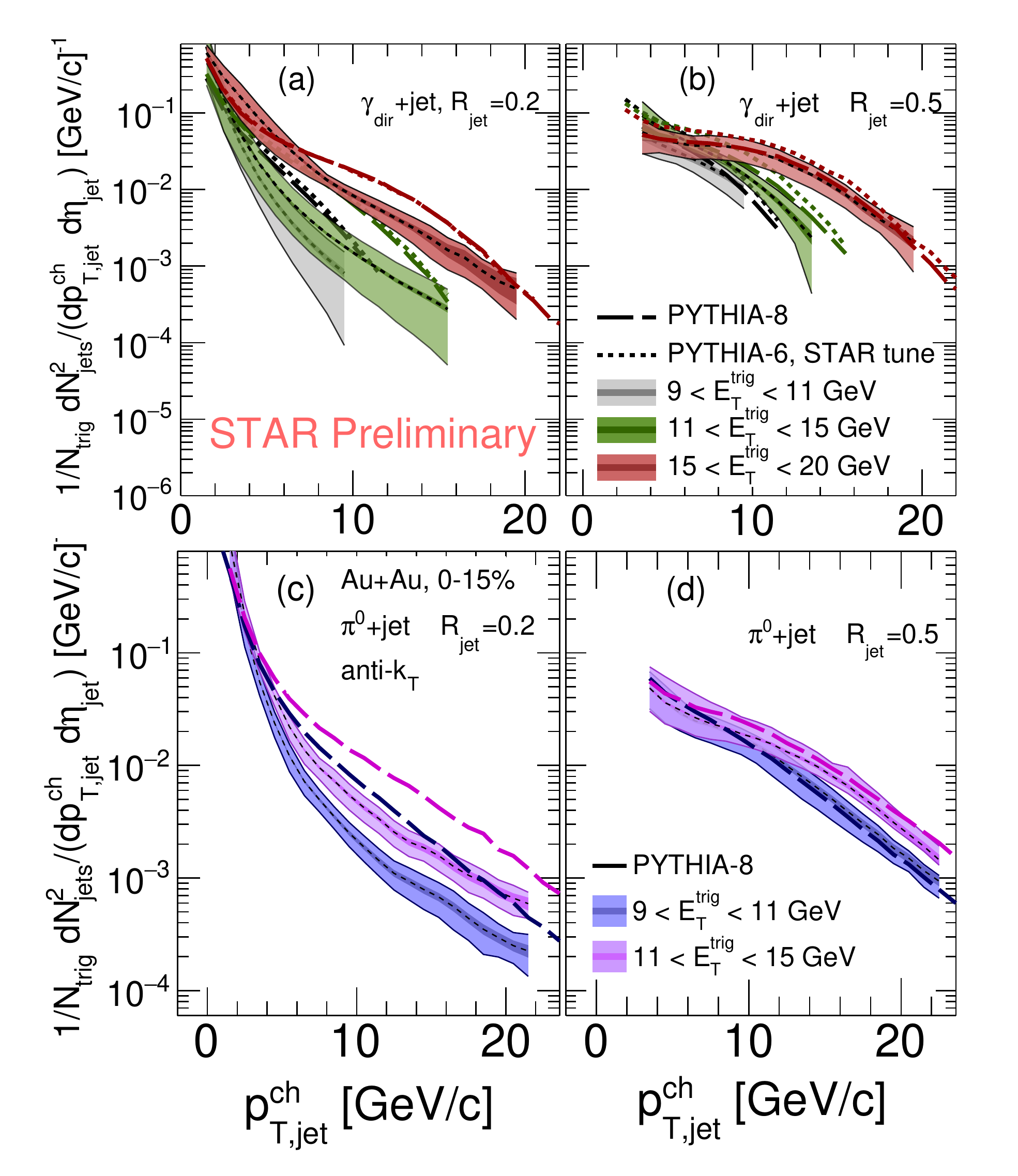}  

 \caption{Semi-inclusive distributions of charged jets recoiling from \DirPho\ (upper) and \piZro\ (lower) triggers. Light and dark bands represent systematic and statistical uncertainties, respectively. Broken and dotted lines represent calculations based on PYTHIA-8 and PYTHIA-6 STAR tune.}
\label{Fig:pTspct}
 \end{wrapfigure}
The offline analysis selects events corresponding to the 0-15\% most central Au+Au collisions, based on uncorrected charged-particle multiplicity within |$\eta$| < 1. The BEMC Shower Max Detector (BSMD) was used offline to select clusters in the range 9 < \ET\ < 20 GeV that have an enhanced population of direct photons (\gammaRich) or \piZro\ ($\pi^{0}_{\rm rich}$). 
A Transverse Shower Profile (TSP) method is used to discriminate between $\pi^{0}_{\rm rich}$~and \gammaRich~triggers~\cite{STAR:2016jdz}. The purity of direct photons in the \gammaRich\ sample is 65--85\% in the range 9 < \ET\ < 20 GeV.  The final corrections are applied on both \gammaRich\ and $\pi^{0}_{\rm rich}$ to get the fully corrected recoil jet yields.  Charged jets are reconstructed with the anti-$k_{\mathrm{T}}$ algorithm ~\cite{Cacciari:2008gp, Cacciari:2011ma} for \R~=~0.2 and 0.5, using charged particle tracks measured in  the Time Projection Chamber (TPC) with $0.2 < \pT < 30$ GeV/$c$ and |$\eta$| < 1. The jet acceptance is |$\eta_{\mathrm{jet}}$| < 1-\R. 

Recoil jets are selected with a $\Delta\phi \in [3\pi/4, 5\pi/4]$, where $\Delta\phi$ is the azimuthal angle between the trigger cluster and the jet axis.
The semi-inclusive distribution is defined as the yield of recoil jets in a bin of transverse momentum ( \pTjet\ ) normalized by the number of triggers. 
The uncorrelated background jet yield in this distribution is corrected using the Mixed-Event (ME) technique developed in~\cite{Adamczyk:2017yhe}. 
Corrections to the recoil jet distributions for instrumental effects and residual \pTjet\ fluctuations due to background are carried out using unfolding methods. The main systematic uncertainties arise from unfolding, ME normalization, and \DirPho~purity.

Due to limited trigger statistics in the current analysis of STAR $pp$ data, the reference distribution from $pp$ collisions is calculated using the PYTHIA event generators. For \DirPho-triggered distributions, both PYTHIA-8 ~\cite{Sjostrand:2007gs} and PYTHIA-6 STAR tune ~\cite{Adam:2019aml} events are used, whereas for \piZro-triggered distributions only PYTHIA-8 is used.

\begin{figure}[htbp]
 \centering
 		\includegraphics[width=0.35\textwidth]{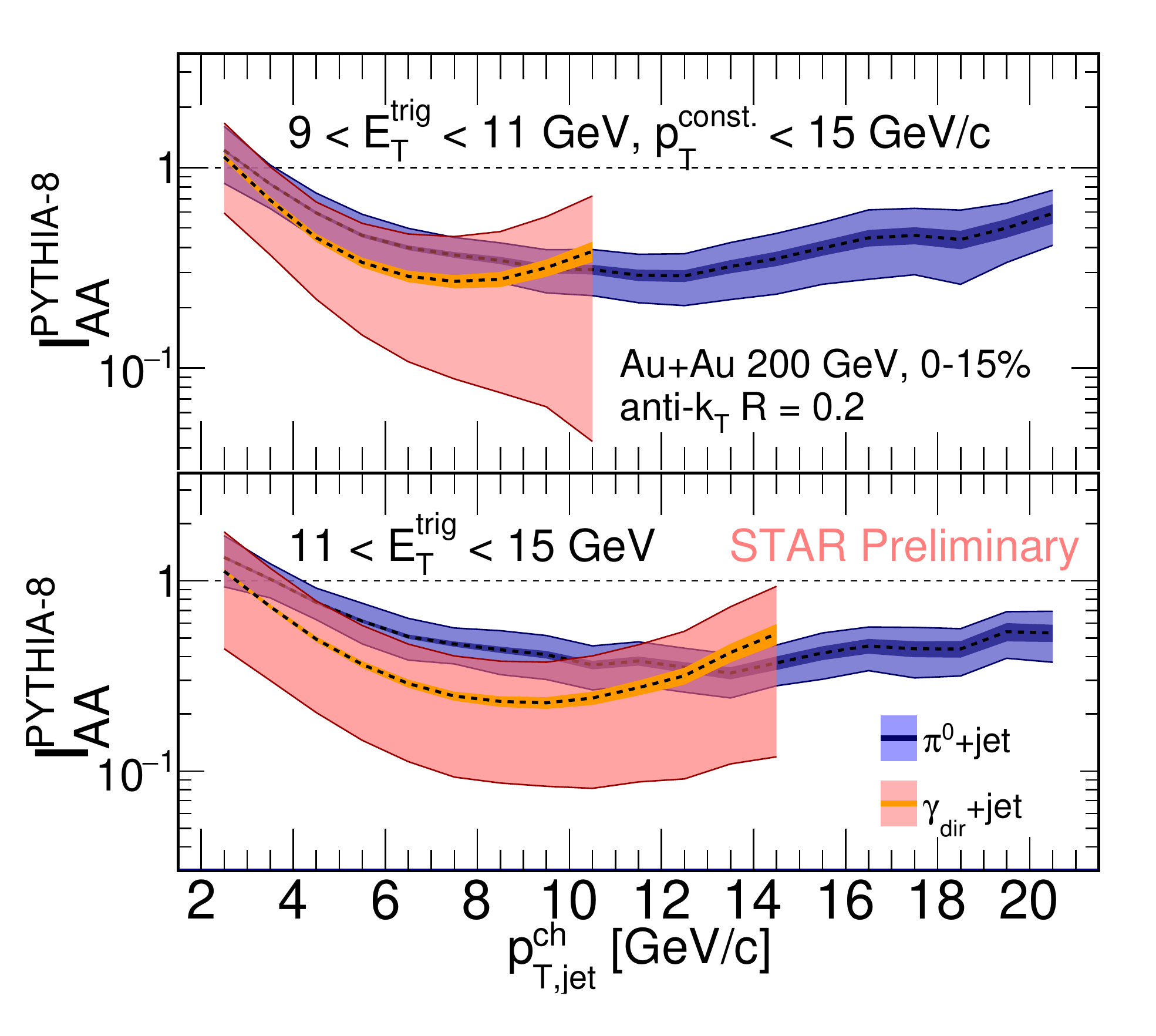}	
		 		\includegraphics[width=0.35\textwidth]{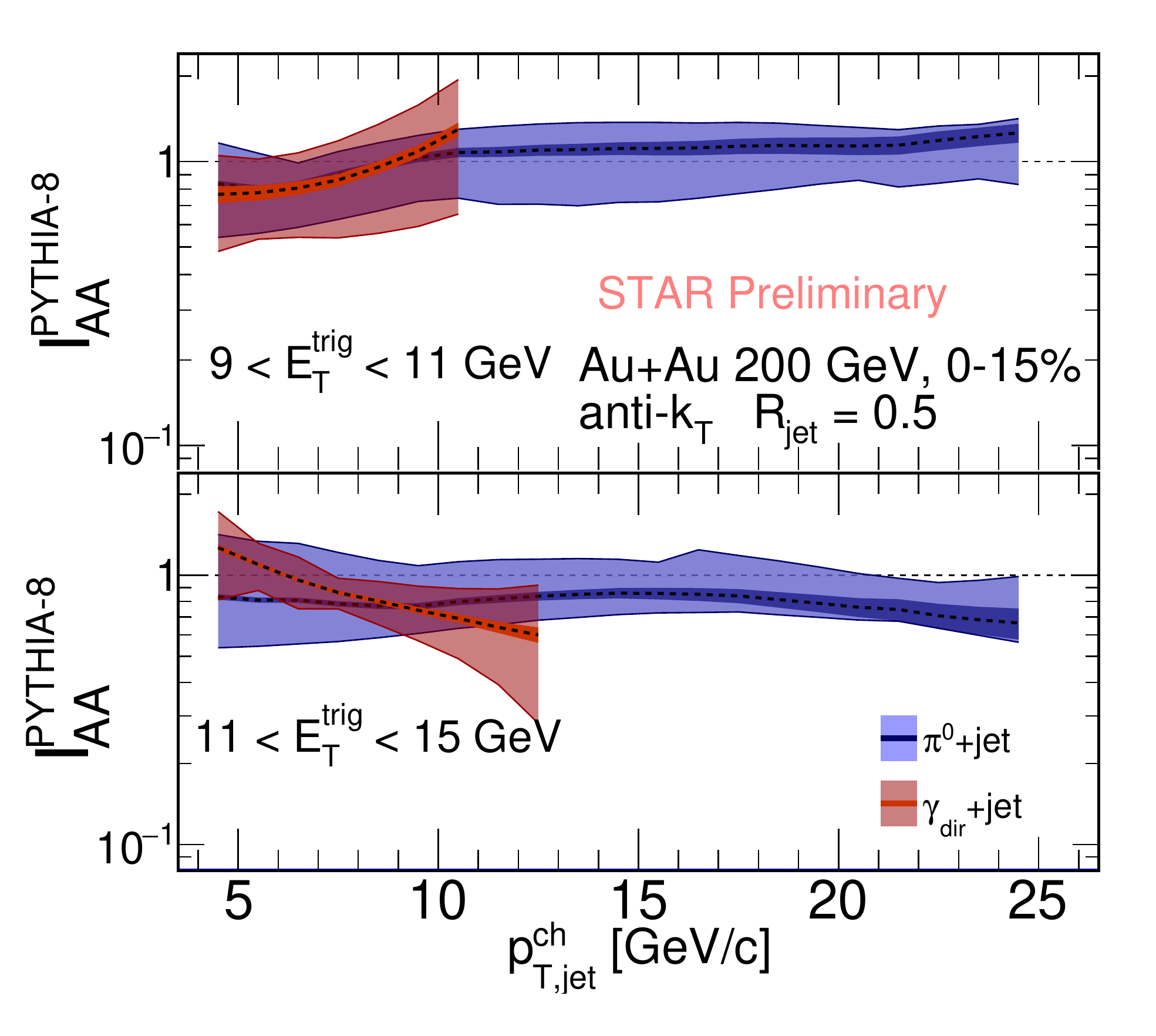}		
\caption{\IAAPYTHIAEIGHT~ vs. \pTjet~for \DirPho\ triggers (red) and \piZro\ triggers (blue) with 9 < \ET<11 GeV (upper) and 11 < \ET < 15 GeV (lower) and for jets with \R~=~0.2 (left) and 0.5 (right). Light and dark bands represent systematic and statistical uncertainties. }
  \label{Fig:IAARecoilJetpT9to15}
\end{figure}

\begin{figure}[htbp]
 \centering
  \includegraphics[width=0.7\textwidth]{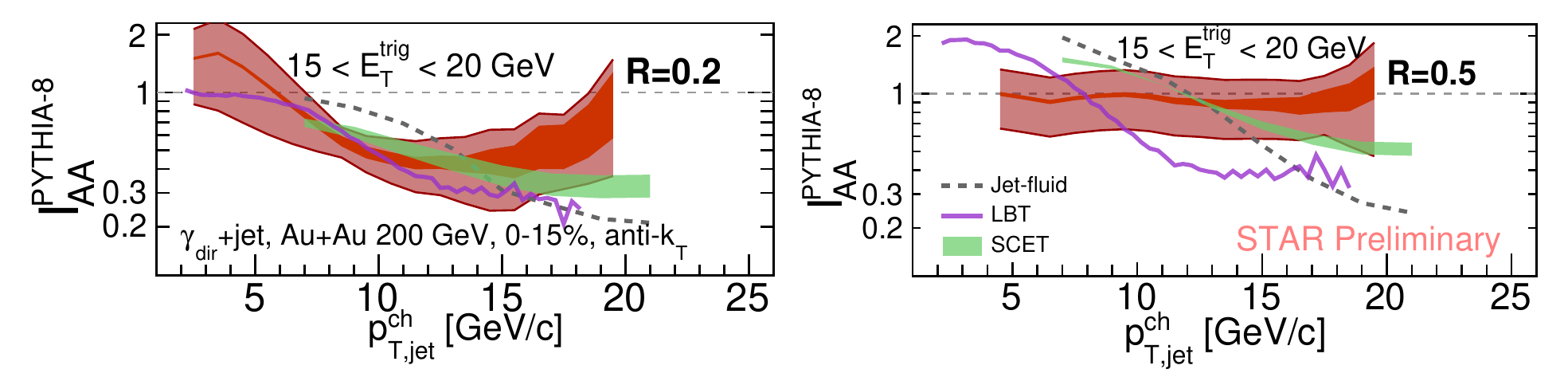}  	  	 
	\includegraphics[width=0.7\textwidth]{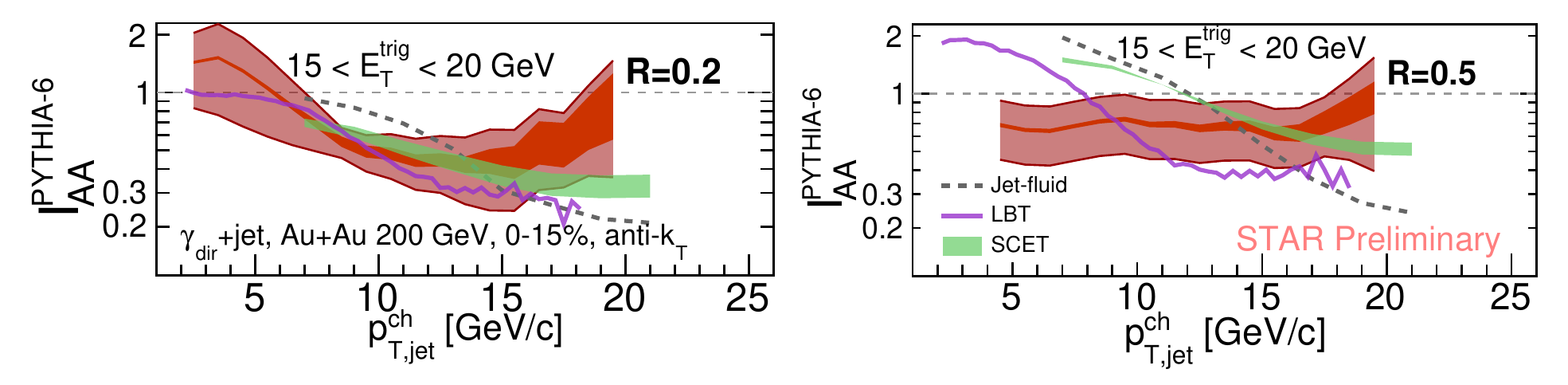}
\caption{
\DirPho+jet: \IAAPYTHIAEIGHT~(upper) and \IAAPYTHIASIX~(lower) vs. \pTjet~for 15 < \ET < 20 GeV  and jets with \R~=~0.2 (left) and 0.5 (right). Light and dark bands represent systematic and statistical uncertainties.
Theory calculations: Jet-fluid~\cite{Chang:2016gjp}, LBT~\cite{Luo:2018pto}, and SCET~\cite{Sievert:2019cwq}.}
  \label{Fig:IAARecoilJetpT15to20}
\end{figure}

\begin{figure}[htbp]
  \centering
    \includegraphics[width=0.36\textwidth]{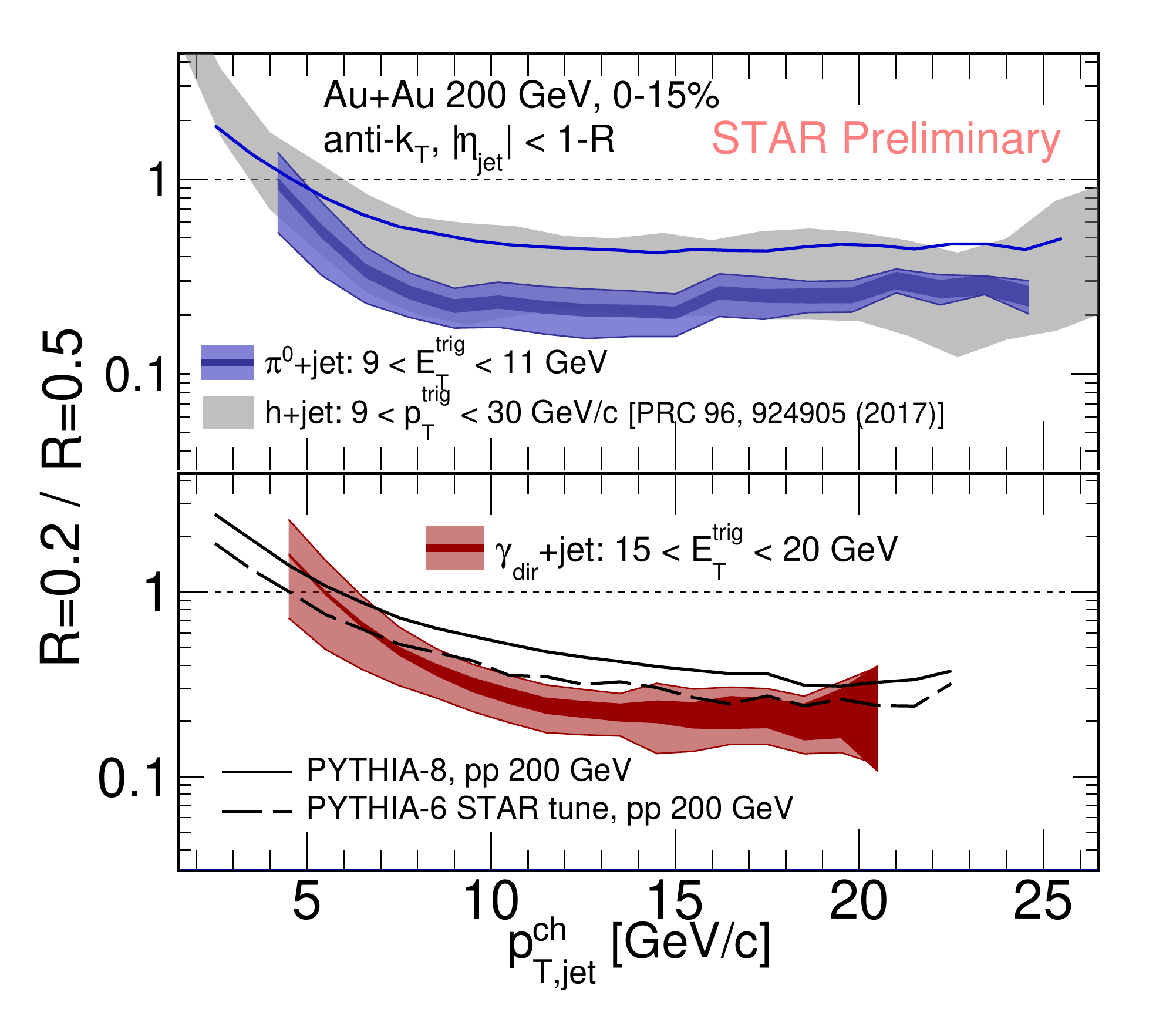}  	  
      \includegraphics[width=0.60\textwidth]{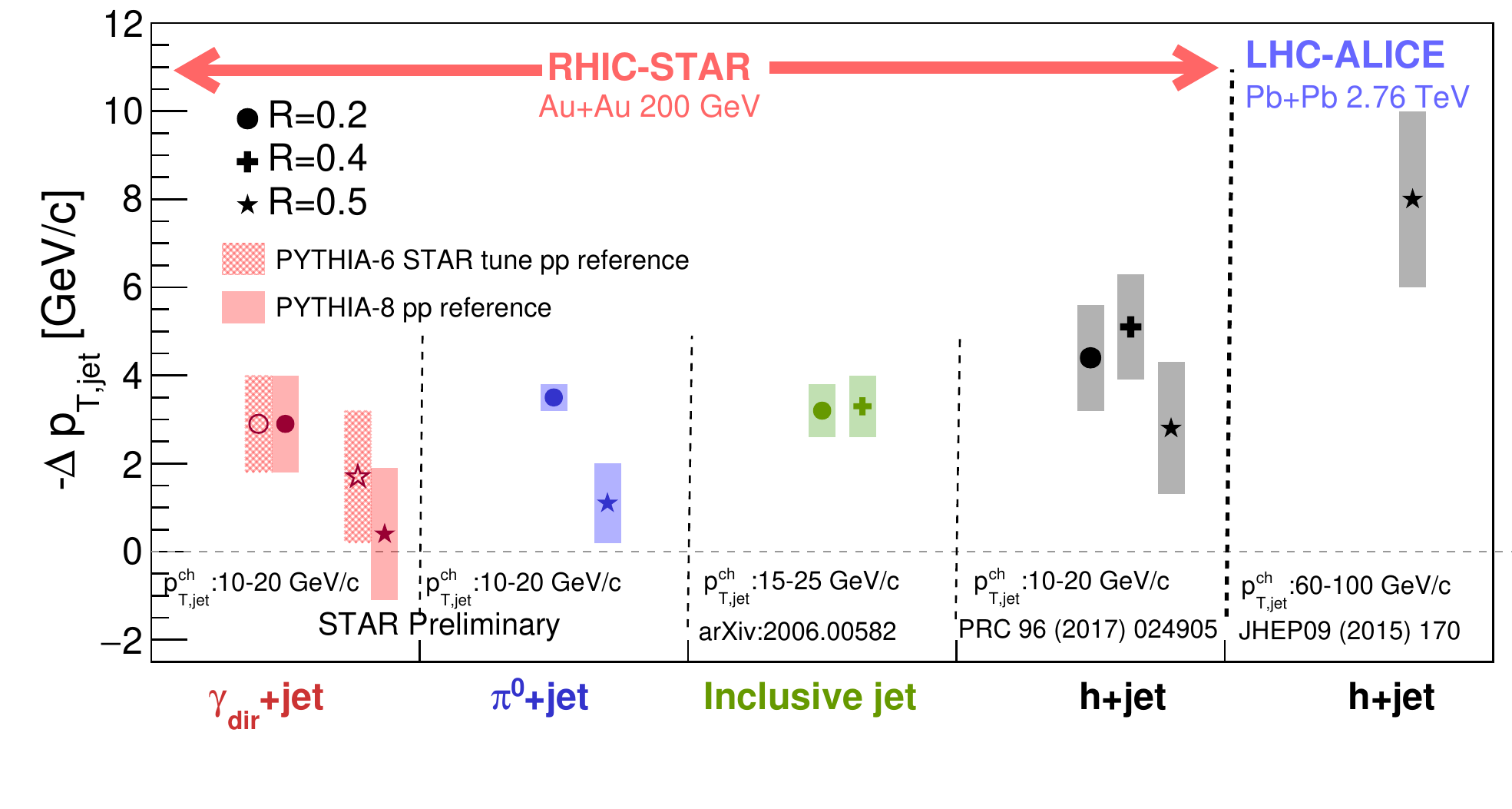}  	  	 	
\caption{Left panel: Ratio of recoil jet yields for \R~=~0.2 and 0.5 as a function \pTjet. Upper: h+jet  and \piZro+jet. Lower: \DirPho+jet. Right panel: The \pTjet~shift (-$\Delta$ \pTjet) for \DirPho+jet, \piZro+jet, inclusive jet, h+jet measurements at RHIC, and h+jet at the LHC.  Note the different \pTjet~ranges.}
  \label{Fig:pTsift}
\end{figure}
 
Figure~\ref{Fig:pTspct} shows fully corrected charged-jet \pT~spectra for \R~=~0.2 and 0.5 recoiling from \DirPho\ in three \ET\ bins, and \piZro\ in two \ET\ bins, measured in central Au+Au collisions and compared to those calculated by PYTHIA for $pp$ collisions. The two PYTHIA versions exhibit negligible difference for \R~=~0.2 and up to 40\% difference for \R~=~0.5. The ratio of recoil jet yield measured in Au+Au collisions to PYTHIA  calculations for pp collisions are denoted as \IAAPYTHIASIX~and \IAAPYTHIAEIGHT\ for the two versions of PYTHIA used.

Figure~\ref{Fig:IAARecoilJetpT9to15} shows \IAAPYTHIAEIGHT~for \DirPho\ and \piZro\ triggers in 9 < \ET\ < 15 GeV for \R~=~0.2 and 0.5. The recoil jet yields show similar suppression for both triggers for \R~=~0.2, with no significant \ET\ dependence. Smaller suppression is observed for \R~=~0.5 for both triggers compared to \R\ = 0.2. 

Figure~\ref{Fig:IAARecoilJetpT15to20} compares \IAAPYTHIAEIGHT\ and \IAAPYTHIASIX\ for \DirPho\ triggers with 15 < \ET < 20 GeV. Comparison is also made to theoretical model calculations~\cite{Chang:2016gjp,Luo:2018pto,Sievert:2019cwq}, which predict different \pT\ dependence to those observed in data.
 
Figure~\ref{Fig:pTsift}, left panel, shows the ratio of recoil jet yields for \R~=~0.2 and 0.5 measured in central Au+Au collisions with both \DirPho\ and \piZro\ triggers. This ratio is sensitive to the jet transverse profile~\cite{Adam:2015doa,Adamczyk:2017yhe}. The \DirPho-triggered ratio is consistent with a calculation based on the PYTHIA-6 STAR tune, indicating no significant in-medium broadening of recoil jets whereas a notable quantitative difference is observed between Au+Au and PYTHIA-8. The ratios for \piZro\ and charged-hadron triggers measured in central Au+Au collisions are consistent within uncertainties.

Jet quenching is commonly measured by yield suppression at fixed \pT\ (\RAA\ and \IAA). However, these ratio observables convolute the effect of energy loss with the shape of the spectrum. To isolate the effect of energy loss alone we convert the suppression to a \pT-shift, -\DeltaJetPt, enabling quantitative comparison of jet quenching measurements with different observables, and comparison of jet quenching at RHIC and the LHC. Figure~\ref{Fig:pTsift}, right panel, shows -\DeltaJetPt\ from this measurement, compared to those of inclusive jets and h+jet at RHIC, and h+jet at the LHC~\cite{Adamczyk:2017yhe,Adam:2020wen,LicenikHP,Adam:2015doa}. The energy loss from the RHIC measurements is largely consistent for the different observables, with some indication of smaller energy loss for \R~=~0.5 than for \R~=~0.2 considering PYTHIA-8 for the vacuum expectation. 
In addition, the results from \R~=~0.2 measurements at RHIC are comparable to those from inclusive \piZro~\cite{Adare:2012wg}. 
An indication of smaller in-medium energy loss is observed at RHIC than at the LHC.

 In summary, we have presented the analysis of semi-inclusive charged-jet distributions recoiling from \DirPho\ and \piZro\ triggers in central Au+Au collisions at \sNN\ = 200 GeV. Significant yield suppression is observed for recoil jets with \R~=~0.2, and a less suppression is seen for \R~=~0.5 using PYTHIA-8 as $pp$ reference. However, the difference between PYTHIA-8 and PYTHIA-6 precludes quantitative conclusions. On the other hand, a definitive conclusion on in-medium jet broadening from the ratio of recoil jet yields at different R can be drawn when the vacuum reference will be resolved by the same measurements in $pp$ collisions at 200 GeV, currently in progress. Theoretical calculations of jet quenching predict a different \pT-dependence of the suppression than that observed in data. Conversion of the measured suppression to a \pT-shift reveals similar energy loss due to the quenching of various jet measurements at RHIC and an indication of smaller energy loss at RHIC than at the LHC.
 
{\bf Acknowledgments:~}This work was supported by the Fundamental Research Funds of Shandong University and DOE DE-SC0015636.

 

\end{document}